\documentclass[12pt]{article}
\usepackage{amssymb,latexsym}
\usepackage[mathscr]{eucal}
\usepackage{amsmath,amssymb}
\usepackage{latexsym}
\usepackage{bbm}
\usepackage{bm}
\newcommand{\beq}{\begin{equation}}
\newcommand{\eeq}{\end{equation}}
\newcommand{\beqn}{\begin{eqnarray}}
\newcommand{\eeqn}{\end{eqnarray}}
\newcommand{\dalpha}{\dot{\alpha}}
\newcommand{\dbeta}{\dot{\beta}}
\newcommand{\nn}{\nonumber}

\numberwithin{equation}{section}

\newcommand{\R}{\mathbbm{R}}
\newcommand{\Z}{\mathbbm{Z}}
\newcommand{\C}{\mathbbm{C}}
\newcommand{\be}{\begin{equation}}
\newcommand{\ee}{\end{equation}}
\newcommand{\ba}{\begin{eqnarray}}
\newcommand{\ea}{\end{eqnarray}}
\newcommand{\bdm}{\begin{displaymath}}
\newcommand{\edm}{\end{displaymath}}

\newcommand{\one}{\mathbbm{1}}

\def\k{\kappa}

\newcommand{\ie}{{\it i.e.\ }}

\DeclareMathAlphabet{\mathpzc}{OT1}{pzc}{m}{it}
\def\bea{\begin{eqnarray}}
\def\eea{\end{eqnarray}}
\def\beas{\begin{eqnarray*}}
\def\eeas{\end{eqnarray*}}
\def\sla{\raise.15ex\hbox{$/$}\kern-.57em}

\def\bea{\begin{eqnarray}}
\def\eea{\end{eqnarray}}

\def\sla{\raise.15ex\hbox{$/$}\kern-.57em}
\def\ie{{\it i.e.}~}

\def\ap{{\alpha^\prime}}

\def\k{\kappa}

\def\vf{\varphi}

\def\cA{{\cal A}}

\def\cC{{\cal C}}
\def\cD{{\cal D}}

\def\cM{{\cal M}}
\def\cN{{\cal N}}

\def\cT{{\cal T}}
\def\cU{{\cal U}}

\def\cZ{{\cal Z}}

\newcommand{\ft}[2]{{\textstyle\frac{#1}{#2}}}
\usepackage[vcentermath,enableskew]{youngtab}
\newcommand{\tinyyoung}[1]{\mbox{\tiny\young(#1)}}

\begin{document}
%%%%%%%%%%%%%%%%%%%%%%%  FRONTESPIZIO  %%%%%%%%%%%%%%%%%%%%%%%%%%%%%%%%%%%
\begin{titlepage}
\begin{flushright}
{ROM2F/2007/09}\\
\end{flushright}
%%%%%%%%%%%%%%%%%%%%%%%  TITOLO  %%%%%%%%%%%%%%%%%%%%%%%%%%%%%%%%%%%%%%%%%
\begin{center}
{\large \sc  D-brane Instantons on the $T^6/\Z_3$ orientifold}\\
\vspace{1.0cm}
%%%%%%%%%%%%%%%%%%%%%%%%%  AUTORI  %%%%%%%%%%%%%%%%%%%%%%%%%%%%%%%%%%%%%%%%%
{\bf Massimo Bianchi}, {\bf Francesco Fucito} and {\bf Jose F. Morales}\\
{\sl Dipartimento di Fisica, Universit\'a di Roma ``Tor Vergata''\\
 I.N.F.N. Sezione di Roma II\\
Via della Ricerca Scientifica, 00133 Roma, Italy}\\
\end{center}
\vskip 2.0cm
%%%%%%%%%%%%%%%%%%%%%%%%%  ABSTRACT  %%%%%%%%%%%%%%%%%%%%%%%%%%%%%%%%%%%%%%
\begin{center}
{\large \bf Abstract}
\end{center}

We give a detailed microscopic derivation of gauge and stringy
instanton generated superpotentials for gauge theories living on
D3-branes at $\Z_3$-orientifold singularities. Gauge instantons
are generated by D(-1)-branes and lead to Affleck, Dine and
Seiberg (ADS) like superpotentials in the effective ${\cal N}=1$
gauge theories with three generations of bifundamental and
anti/symmetric matter. Stringy instanton effects are generated by
Euclidean ED3-branes wrapping four-cycles on $T^6/\Z_3$. They give
rise to Majorana masses, Yukawa couplings or  non-renormalizable
superpotentials depending on the gauge theory. Finally we determine the
conditions under which ADS like superpotentials are generated in
${\cal N}=1$ gauge theories with adjoints, fundamentals, symmetric
and antisymmetric chiral matter.

\vfill
\end{titlepage}

\tableofcontents

\section{Introduction}

Our understanding of non perturbative effects in  four dimensional
supersymmetric gauge theories (SYM) has dramatically improved in
recent years. This is due mainly to the observation that integrals
over the moduli space of gauge connections localize around a
finite number of points\cite{Nekrasov:2002qd}. These techniques
have been applied to the study of multi-instanton corrections to
${\cal N}=1, 2, 4$ supersymmetric
 gauge theories in $\R^4$ \cite{Flume:2002az,Bruzzo:2002xf,Losev:2003py,Nekrasov:2003rj,Flume:2004rp,Marino:2004cn,Fucito:2004gi,Fucito:2005wc,Fujii:2007qe}
 (see \cite{Dorey:2002ik,Bianchi:2007ft} for reviews of multi-instanton techniques
 before localization and complete lists of references).
 In the
D-brane language language, the dynamics of the
gauge theory around the instanton background is described by
an effective theory  governing the interactions of the lowest energy excitations
of open strings ending on a  bound state of Dp-D(p+4) branes. For the case of ${\cal
N}=2, 4$ SYM the multi-instanton action has been derived via
string techniques in \cite{Billo:2002hm,Billo:2006jm}.

In \cite{Akerblom:2006hx}, D-brane techniques have been applied to
the computation of the Affleck, Dine and Seiberg (ADS)
superpotential \cite{Affleck:1983rr,Affleck:1983mk} for ${\cal
N}=1$ SQCD with gauge group $SU(N_c)$ and $N_f=N_c - 1$ massless
flavours and $Sp(2N_c)$ with $2N_f = 2N_c$ flavours.
The ${\cal N}=1$
gauge theory is realized on the four-dimensional intersection of
$N_c$ coloured and $N_f$ flavour D6 branes. Chiral matter comes
from strings connecting the flavor and color D6 branes.
  Instantons in the $U(N_c)$ gauge theory are realized in terms of ED2 branes
parallel to the stack of $N_c$ D6-branes. By careful integrating
the supermoduli (massless strings with at least one end on the
ED2) the precise form of the ADS superpotential was reproduced in
the low energy, field theory limit $\ap \rightarrow 0$. In the
recent literature ED2-brane instantons in intersecting D6-brane
models have received particular attention in connection with the
possibility of generating a Majorana mass for right handed
neutrinos and their superpartners
\cite{Blumenhagen:2006xt,Haack:2006cy,Ibanez:2006da,Florea:2006si,Cvetic:2007ku}.
The field theory interpretation of this new instanton effect is
far from clear and it is the subject of active investigation.
 In this paper we present a detailed derivation of these new non
perturbative superpotentials in ${\cal N}=1$ $\Z_3$-orientifold models.
Investigations of stringy instantons on ${\cal N}=1$ $\Z_2\times \Z_2$
orientifold singularities appeared recently in \cite{Argurio:2007}.

We study SYM gauge theories living on D3 branes located at a
$\Z_3$-orientifold singularity. There are two choices for the
orientifold projection \cite{Sagnotti:1987tw, Pradisi:1988xd,
Bianchi:1990yu,Bianchi:1990tb, Bianchi:1991eu} realized by two
types of O3-planes\footnote{We will only consider O3$^\pm$-planes,
not the more exotic $\widetilde{\rm O3}^\pm$-planes
\cite{Witten:1997bs,Dudas:2000bn,Angelantonj:2002ct}.}. They lead
to anomaly free\footnote{Factorizable $U(1)$ anomalies are
cancelled by a generalization of the Green-Schwarz mechanism
\cite{Green:1984sg, Sagnotti:1992qw, Ibanez:1998qp,
Bianchi:2000de, Antoniadis:2002cs, Anastasopoulos:2003aj,
Anastasopoulos:2004ga, Anastasopoulos:2003aj} that may require the
introduction of generalized Chern-Simons couplings
\cite{Anastasopoulos:2006cz}.} chiral ${\cal N}=1$  gauge theories
with gauge groups $SO(N-4)\times U(N)$ or $Sp(N+4)\times U(N)$ and
three generations of chiral matter in the bifundamental and
anti/symmetric representation of $U(N)$. The archetype of this
class can be realized as a stack of $3 N + 4$ D3-branes and one
O3$^-$-plane sitting on top of an $\R^6/\Z_3$ singularity. This
system can be thought of as a T-dual local description (near the
origin) of the $T^6/\Z_3$ type I string vacuum found in
\cite{Angelantonj:1996uy}. The lowest choices of $N$ lead to
$U(4)$ or $U(5)$ gauge theories with three generations of chiral
matter in the ${\bf 6}$ and ${\bf 10}+{\bf 5}^*$ that are
clearly of phenomenological interest in unification scenarios
\cite{Uranga:2003pz, Kiritsis:2003mc, Blumenhagen:2005mu,
Blumenhagen:2006ci} \footnote{Only the $U(4)$ case can be realized in the compact
$\Z_3$ orientifold. In general the Chan-Paton group is
$SO(8-2n)\times U(12-2n) \times H_n$ where $H_n = U(n)^3, SO(2n),
U(n), U(1)^n$ depending on the choice of Wilson lines
\cite{Bianchi:1991eu, Bianchi:1997rf, Cvetic:1999qx,
Cvetic:2000aq, Cvetic:2000st}.}.

In \cite{Bianchi:2007fx} the $U(4)$ case was studied  and the form
of the ADS-like superpotential was determined combining
holomorphicity, $U(1)$ anomaly, dimensional analysis and flavour
symmetry.  Stringy instanton effects were also considered. Very
much as for worldsheet instantons in heterotic strings
\cite{Dine:1986zy, Dine:1987bq, Witten:1999eg, Beasley:2003fx,
Beasley:2005iu}, these genuinely stringy instantons give rise to
superpotentials that do not vanish at large VEV's of the open
string (charged) `moduli'.

Here we derive the non-perturbative superpotentials from a direct
integration over the D-instanton super-moduli space. Gauge instantons
are described in terms of open strings ending on D(-1) branes while
stringy instantons are given by open strings ending on euclidean ED3
branes wrapping a four cycle inside the Calabi Yau .
The open strings connecting the stack of D3 branes to D(-1) and ED3 branes
have four and eight mixed Neumann-Dirichelet directions respectively.
This ensures that the bound state is supersymmetric.
The superpotential receives contribution
from disk, one-loop annulus  and M\"obius
 amplitudes ending on the D(-1) or ED3 branes. We find that ADS superpotentials
 are generated only for  two gauge theory choices $U(4)$ and $Sp(6)\times U(2)$
 inside the $\Z_3$-orientifold class. Stringy instantons leads to Majorana masses in
 the U(4) case, Yukawa couplings in the $U(6)\times SO(2)$ gauge theory and
 non-renormalizable couplings for $U(2N+4)\times SO(2N)$ gauge theories with $N>3$.

The plan of the paper is as follows.

In section \ref{gaugeth} we  review the gauge theories coming
from a stack of D3 branes at a  $\C^3/\Z_3$ orientifold singularity.
In  Section \ref{sd1instanton}
we consider non-perturbative effects generated by D(-1) gauge instantons,
corresponding to ADS-like superpotential  in the low energy limit. A detailed
analysis of one-loop vacuum amplitudes
 and the integrals over the supermoduli is presented for SYM theories
   with gauge groups $Sp(6)\times U(2)$
and $U(4)$.
In section \ref{newinst}, we consider stringy instanton
effects generated by ED3-branes.  Once again a detailed analysis of the
the  one-loop string amplitudes and the integrals over the supermoduli is presented.
In section \ref{sads} we present a ``complete'' list of ${\cal N}=1$ SYM theories with
matter in the adjoint, fundamental, symmetric and antisymmetric representation of the gauge
groups (U, SO, Sp) which exhibit a non perturbatively generated ADS superpotential.

We conclude with some comments and directions for future
investigation in Section \ref{conclusions}.

\section{The Gauge Theory}
\label{gaugeth}

The low energy dynamics of the open strings living on a stack of N
D3-branes in flat space is described by a ${\cal N}=4$ $U(N)$ SYM
gauge theory. In the  ${\cal N}=1$ language the fields are grouped
into a vector multiplet
$V=(A_\mu,\lambda_\alpha,\bar{\lambda}_{\dalpha})$ and 3 chiral
multiplets $\Phi^I=(\phi^I,\psi^I_\alpha)$, $I=1,2,3$ , all in the
adjoint of the gauge group.

We consider the D3-brane system at a $\R^6/\Z_3$ singularity. At
the singularity the $N$ D3-branes group into stacks of $N_n$
fractional branes with $n=0,1,2$ labelling the conjugacy classes
of $\Z_3$. The gauge group $U(N)$ decomposes as $\prod_n
U(N_n)$.
 More precisely, denoting by $\gamma_{_{\theta,N}}$ the projective embedding of the
orbifold group element $\theta\in\Z_3$ in the Chan-Paton group and imposing
$\gamma_{_{\theta,N}}^3= 1$ and $\gamma_{_{\theta,N}}^\dagger =
\gamma_{_{\theta,N}}^{-1}$ one can write
 \be
 \gamma_{_{\theta^h,N}} = ({\bf 1}_{_{N_0\times \bar{N}_0}},
\omega^h \,{\bf 1}_{_{N_1\times \bar{N}_1}}, \bar\omega^h\,{\bf 1}_{_{
N_2\times \bar{N}_2}})
\label{gz3}
\ee
 with $N=\sum_n N_n$.
 The resulting gauge theory can be found by projecting the ${\cal N}=4$ $U(N)$
 gauge theory under the $\Z_3$ orbifold group action:
\beq V\to \gamma_{_{\theta,N}}\, V\, \gamma_{_{\theta,N}}^{-1} \quad\quad
\Phi^I \to \omega\, \gamma_{_{\theta,N}}\,\Phi^I \,
\gamma_{_{\theta,N}}^{-1}\quad\quad \omega=e^{2\pi i/3}
\label{z3action} \eeq
 Keeping only invariant components under (\ref{z3action}) one finds the ${\cal N}=1$  quiver gauge theory
\beqn
V : &&{\bf N_0 \bar{N}_0}+{\bf N_1 \bar{N}_1}+{\bf N_2 \bar{N}_2}\nn\\
\Phi^I :&& 3\times \left[{\bf N_0 \bar{N}_1}+{\bf N_1
\bar{N}_2}+{\bf N_2 \bar{N}_0} \right] \label{quiver} \eeqn with
gauge group $\prod_n U(N_n)$ and three generations of
bifundamentals. More precisely $V$ and $\Phi^I$ are $N\times N$ block matrices ($N=\sum_n N_n$)
with non trivial $N_n\times \bar{N}_m$ blocks given by (\ref{quiver}). Under $\Z_3$
a block $N_n\times \bar{N}_m$ transform as $\omega^{n-m}$.
These non-trivial transformation properties are compensated by the space-time eigenvalues of the
corresponding field ($\omega^0$ for $V$ and $\omega$ for $\Phi^I$ )
making the corresponding component invariant under $\Z_3$.

Next we consider the effect of introducing an O3$^{\pm}$-plane.
Woldsheet parity $\Omega$ flips open string orientations and act on
Chan-Paton indices as $N_n\leftrightarrow \bar N_{-n}$ where subscripts are
always understood mod 3. This prescription leads to
\beq
\Omega:\quad\quad N_0\leftrightarrow \bar{N}_0 \quad\quad N_1 \leftrightarrow \bar{N}_2
\label{omegapresc}
\eeq
The choices of O3$^\pm$-planes correspond to keep
states with eigenvalues $\Omega=\pm 1$  and lead to symplectic or
orthogonal gauge groups\footnote{In the compact case, realized in
terms of D9-branes and O9-plane on $T^6/\Z_3$, the orthogonal
choice is dictated by global tadpole cancellation. Turning on a
quantized NS-NS antisymmetric tensor \cite{Bianchi:1991eu,Bianchi:1997rf,Witten:1997bs} leads to
symplectic groups.}.

We start by considering the O3$^-$ case.
  Keeping $\Omega=-$ components from (\ref{quiver}) one finds
\beqn
V : &&{\bf \ft12 N_0(N_0-1)}+{\bf N_1 \bar{N}_1}\nn\\
\Phi^I :&& 3\times \left[
 {\bf N_0 \bar{N}_1}+{\bf \ft12 N_1(N_1-1)} \right]
\label{mattercontent}
\eeqn
This follows from (\ref{quiver}) after identifying the mirror images $\bar N_0= N_0$, $\bar N_2= N_1$,
and antisymmetrizing the resulting block matrix.
(\ref{mattercontent}) describes the field content of a ${\cal N}=1$ SYM
 with gauge group $SO(N_0)\times U(N_1)$ and three chiral  multiplets in the
$\left[(\tinyyoung{\hfil},\bar{\tinyyoung{\hfil}})+(\bullet,\tinyyoung{\hfil,\hfil})\right]$.

For general $N_0,N_1$ the $U(N_1)$ gauge theory is anomalous. The
anomaly is a signal of the presence of a twisted RR tadpole
\cite{Ibanez:1998qp,Bianchi:2000de}.
Focusing on a local description near the
orientifold singularity one can relax the global tadpole
cancellation  condition \cite{Aldazabal:2000sa,Buican:2006sn}.
These models can be thought as local descriptions of a more complicated Calabi Yau
near a $\Z_3$ sigularity.
Cancellation of the twisted RR tadpole can be
written as \cite{Angelantonj:1996uy}
\beq
{\rm tr}\,\gamma_{_{\theta,N}} =-4 \quad\quad \Rightarrow \quad\quad N_0=N_1-4
\label{anomalycanc}\eeq
and ensures the cancellation of the irreducible
four-dimensional anomaly
\beq
I(F)\sim \left[ -N_0+(N_1-4)\right]
{\rm tr} \,F^3=0
\eeq
Finally the running of the gauge coupling constants is governed by
the $\beta$ functions with one-loop coefficients\footnote{
Here ${\rm tr}_{\bf \cal R} T^a T^b =\ell({\bf \cal R})$, i.e.
$\ell({\bf N})=\ft12$, $\ell({\bf N \bar{N}})= N$ and $\ell({\bf \ft12N(N\pm 1)})=\ft12(N\pm 2)$.}
\bea
\beta_0 &=& 3\, \ell({\bf \ft12 N_0(N_0-1)})- 3 N_1\,\ell( {\bf N_0}) \nn\\
&=& \ft32 (N_0-N_1-2)=-9
\quad\quad \rm{(IR \ free)}\nn \\
\beta_1 &=& 3\, \ell({\bf  N_1\bar{N}_1})- 3N_0\,\ell( {\bf \bar{N}_1})-
3\, \ell( {\bf  \ft12 N_1(N_1-1)})\nn\\
&=& \ft32 (-N_0+N_1+2)=+9 \quad\quad \rm{(UV \ free)}\label{betaG}
\eea
 with $\beta_n$ refering to the $n^{\rm th}$-gauge group.
The last equalities arise after imposing the anomaly cancellation (\ref{anomalycanc}).
 As expected, $\beta_0+\beta_1=0$  since the ten-dimensional dilaton does not run.

 The case $\Omega=+$ works in a similar way. The resulting ${\cal N}=1$ quiver has
 gauge group $Sp(N_0)\times U(N_1)$ and three chiral multiplets in
 the
 $\left[(\tinyyoung{\hfil},\bar{\tinyyoung{\hfil}})+(\bullet,\tinyyoung{\hfil\hfil})\right]$.
 The $U(N_1)$ is anomaly free for $N_0=N_1+4$ and the one-loop $\beta$ function
 coefficients are given by $\beta_0=+9$ (UV free) and $\beta_1=-9$ (IR free).

\section{D(-1) Instantons}
\label{sd1instanton}

There are two sources of supersymmetric instanton corrections in
the D3 brane gauge theory: D(-1)-instantons and Euclidean ED3-branes
wrapping four cycles on $T^6/\Z_3$. Both are point-like
configurations in the space-time and can be thought of as D(-1)-D3 and ED3-D3
bound states with four and eight directions with mixed Neumann-Dirichlet boundary conditions.

\subsection{D3-D(-1) in flat space}

Gauge instantons in SYM
%\cite{Amati:1988ft,Shifman:1999mv,Ibanez:1998qp, Bianchi:2000de}
can be efficiently described in terms of D(-1)-branes living on the world-volume of
D3-branes \cite{Douglas:1995bn}. As
before, we start from the ${\cal N}=4$ case: a bound state of $N$
%\footnote{Here, we
%will not pursue the holographic viewpoint advocated \eg in
%\cite{Bianchi:1998nk,Fucito:2001ha,Dorey:2002ik}.}
D3 and $k$ D(-1) branes in flat space. In this formalism
instanton moduli are described by the lowest energy modes of open strings with at least
one end on the
D(-1)-brane stack. The gauge theory dynamics around the instanton
background can be described in terms of the $U(k)\times U(N)$
0-dimensional matrix theory living on the D-instanton
world-volume. In particular, the ADHM constraints
\cite{Atiyah:1978ri} defining the moduli space of self-dual YM
connections follow from the F- and D- flatness condition in the
matrix theory \cite{Douglas:1995bn}.

The instanton moduli space is given by the D(-1)D3 field content
 \beqn
(a_{\mu}, \theta^A_{
\alpha} , \chi_{a}, D^c,  \bar{\theta}_{A\dalpha} ) &&  {\bf k}{\bf \bar{k}}\nn\\
(w_{\dalpha},\nu^A) && {\bf k}{\bf \bar{N}}\nn\\
(\bar{w}_{\dalpha}, \bar{\nu}^{A}) &&  {\bf N}{\bf \bar{k}}
\label{d3d10}\eeqn with $\mu=1,\ldots,4$, $\alpha,\dalpha=1,2$
(vector/spinor indices of $SO(4)$), $a=1,\ldots,6$, $A=1,\ldots,4$
(vector/spinor indices of $SO(6)_R$), $c=1,\ldots,3$. The matrices $a_\mu,\chi_a$
describe the positions of the instanton in the directions parallel and perpendicular to the D3-brane
respectively,
$w_\alpha$ is given by the NS open D3-D(-1) string (instanton sizes and orientations), $D^c$ are
auxiliary fields and $\theta^A_{\alpha},\bar{\theta}_{A\dalpha},\nu^A$ are the fermionic
superpartners.

  The D3-D(-1) action
   can be written as \cite{Dorey:1999pd}
\be
S_{k,N}={\rm tr}_{k}\left[{1\over g_0^{2}}S_{G} + S_{K}+S_{D}
\right]\label{cometipare}
\ee
with
\bea
\label{Sd}
&&S_{G}=
-[\chi_a,\chi_b]^2+ i \bar{\theta}_{\dot{\alpha}
A}[\chi_{AB}^\dagger,\bar{\theta}^{\dot{\alpha}}_B] -D^{c}D^{c}
\\
&&S_{K}=  -[\chi_a,a_{\mu}]^2 +\chi_a
\bar{w}^{\dalpha} w_{\dalpha}\chi_a -  i\theta^{\alpha
A}[\chi_{AB} \theta^{B}_{\alpha}]+2  i
\chi_{AB} \bar{\nu}^{A} \nu^{B}  \, \nonumber\\
&&S_{D}=  i \left( -[a_{\alpha\dot{\alpha}},\theta^{\alpha A}]
+\bar{\nu}^{A} w_{\dot{\alpha}}
+\bar{w}_{\dot{\alpha}}\nu^{A}\right)
\bar{\theta}^{\dot{\alpha}}_{A} +D^{c}\left(\bar{w} \sigma^c w-i
\bar{\eta}_{\mu\nu}^c [a^\mu,a^\nu] \right) \nonumber \eea with
$\chi_{AB}\equiv {1\over 2} \cT^a_{AB} \chi_a$,
$\cT^a_{AB}=(\eta^c_{AB},i\bar\eta^c_{AB})$ given in terms of the
t'Hooft symbols and $g^2_0=4\pi(4\pi^2 \alpha')^{-2}\, g_s$.   The action (\ref{Sd}) follows from the
dimensional reduction of the
 D5-D9 action in six dimensions down to zero dimension.  As a consequence
 our subsequent results hold up to some computable non vanishing
 numerical constant.

In the presence of a v.e.v.
for the six $U(N)$ scalars $\varphi_a$ in the  D3-D3 open string sector
we must add to $S_{k,N}$
\be
S_\vf=  {\rm tr}_{k}\left[\bar{w}^{\dalpha}  (\vf_a \vf_a + 2 \chi_a \vf_a) w_{\dalpha}
 +2 i \bar{\nu}^{A} \vf_{AB} \nu^{B}\right]\label{Sd111}
\ee

The multi-instanton partition function is
\beqn \cZ_{k,N} = \int_{\mathfrak{M}}
e^{-S_{k,N}-S_\vf}=\frac{1}{{\rm Vol}\,U(k)}\, \int_{\mathfrak{M}} d\chi\, d D\,da\,
d\theta\,d\bar{\theta}
d w\, d\nu\, e^{-S_{k,N}-S_\vf} \nonumber
\label{partfunc}
\eeqn
 In the limit $g_0\sim (\alpha')^{-1}\rightarrow\infty$, gravity decouples
from the gauge theory and the contributions coming from $S_G$ are
suppressed. The fields
$\bar{\theta}_{\dalpha A}$, $D^c$ become Lagrange multipliers
implementing the super ADHM constraints
\bea
 \bar{\theta}_{\dalpha A}:  &&\bar{\nu}^{A} w_{\dot{\alpha}} +\bar{w}_{\dalpha}\nu^{A}
-[a_{\alpha\dot{\alpha}},\theta^{\alpha A}]=0\nonumber\\
  D^c:  && \bar{w} \sigma^c w-i \bar{\eta}_{\mu\nu}^c [a^\mu,a^\nu]=0
\label{adhm} \
\eea

\subsection{D(-1)-D3 at the $\C^3/\Z_3$-orientifold }

Let us now consider in turn the $\Omega$ and then the
$\Z_3$ projection.

The effect of introducing an O3$^{\pm}$-plane in the D(-1)-D3
system corresponds to keep open string states with
 eigenvalue $\Omega I=\pm$, $\Omega$ being the worldsheet parity and
 $I$ a reflection along the Neumann-Dirichlet  directions of the Dp-O3 system
 \cite{Fucito:2004gi}.
 On D(-1) string modes, $I$ acts as a reflection in the spacetime plane
 \beqn
 I  : &&  a_{\mu}\to -a_{\mu} \quad\quad \theta^A_{\alpha}\to
-\theta^A_{\alpha}\label{iact}
\eeqn
leaving all other moduli invariant.
  In addition consistency with the D3-O3
 projection requires that the D(-1)  strings are projected in the opposite way with respect
to the D3-branes\cite{Gimon:1996rq} . {From} the gauge theory
point of view this corresponds to the well known fact that $SO(N)$
and $Sp(N)$ gauge instantons have ADHM constraints invariant under
$Sp(k)$ and $SO(k)$ respectively.

  We start by considering the O3$^-$ case.
  After the   $\Omega I$ projection the surviving fields are
\beqn
(a_{\mu},  \theta^A_{
\alpha} ) && {\bf \ft12 k(k-1)}  \nn\\
( D^c ,  \chi^I, \bar{\chi}_I,  \bar{\theta}_{A\dalpha} )  &&  {\bf \ft12 k(k+1)} \nn\\
(w_{\dalpha},\nu) && {\bf k  N} \quad . \label{inv0}
\eeqn
Since we are dealing with a $SO(N)$ gauge theory the $D^c$ moduli are projected
in the adjoint of $Sp(k)$. This is also the case for all the other moduli even
under $I$ while the odd ones, $(a_{\mu},  \theta^A_{\alpha} )$, turn out to be antisymmetric.

Let us
now consider the $\Z_3$ projection. Out of the six $\chi_a$ one
can form three complex fields $\chi^I$ with eigenvalues $\omega$
under $\Z_3$ and their conjugate $\bar\chi_I$. To embed the $\Z_3$
projection into $SU(4)$ we decompose the spinor index $A=(0,I)$,
with $I=1,\ldots,3$ and the zeroth direction along the
surviving ${\cal N}=1$ supersymmetry. The D3 and D(-1)
gauge groups $SO(N)$ and $Sp(k)$ break into $SO(N_0)\times
U(N_1)$ and $Sp(k_0)\times U(k_1)$ respectively with $N_0$ ($k_0$) the
number of fractional D3 (D(-1)) branes invariant under $\Z_3$ and
$N_1$ ($k_1$) those transforming with eigenvalue $\omega$. More
precisely, the projective embedding of the $\Z_3$ basic orbifold group
element $\theta$ in the Chan-Paton group can be written
 \bea
 \gamma_{_{\theta^h,N}}  &=&  ({\bf 1}_{_{N_0\times  N_0}},
\omega^h\, {\bf 1}_{_{N_1\times \bar N_1}}, \bar\omega^h\,{\bf 1}_{_{\bar  N_1 \times N_1}}) \nn\\
\gamma_{_{\theta^h,k}}  &=&  ({\bf 1}_{_{k_0\times  k_0}}, \omega^h\,
{\bf 1}_{_{k_1\times \bar k_1}}, \bar\omega^h\,{\bf 1}_{_{ \bar k_1\times  k_1 }}) \eea
After projecting under $\Z_3$  the symmetric/antisymmetric matrices in (\ref{inv0})
break into $k_m\times \bar k_n$, $k_m\times \bar N_n$ or $N_m \times \bar k_n$
each transforming with eigenvalue  $\omega^{m-n}$. In addition fields with up(down)
index $I$ transform like $\omega(\bar{\omega})$.
 Keeping only the invariant components one finds
 \beqn
(a_{\mu} ;  \theta^0_{
\alpha} ) && {\bf \ft12 k_0(k_0-1)}+{\bf k_1 \bar{k}_1} \nn\\
\theta^I_{\alpha} && {\bf \ft12 k_1(k_1-1)}+ {\bf k_0 \bar{k}_1}\nn\\
( D^c ;  \bar{\theta}_{0\dalpha} )  &&  {\bf \ft12 k_0(k_0+1)}+{\bf  k_1 \bar{k}_1}\nn\\
( \bar{\chi}_{I} ;  \bar{\theta}_{I\dalpha} ) &&
{\bf \ft12 \bar{k}_1(\bar{k}_1+1)}+{\bf k_0 k_1}\nn\\
\chi^I &&
{\bf \ft12 k_1(k_1+1)}+ {\bf k_0 \bar{k}_1}\nn\\
(w^{\dalpha} ; \nu^0) && {\bf k_0 N_0}+{\bf  k_1 \bar{N}_1}+ {\bf  \bar{k}_1 N_1}\nn\\
\nu^I &&{\bf k_0 \bar{N}_1}+{\bf  \bar{k}_1 N_0}+ {\bf k_1
N_1}\label{inv} \eeqn
Notice that the $\Z_3$ eigenvalues of the Chan-Paton indices in the r.h.s. of (\ref{inv})
compensate for those of the moduli in the l.h.s.
making the field invariant under $\Z_3$. In addition (odd)even components
under $I$ are (anti)symmetrized ensuring the invariance under $\Omega I$.

The multi-instanton action follows from that of $N=N_0+2N_1$
D3 branes and $k=k_0+2k_1$ D(-1) instanton in flat space
(\ref{cometipare}) with $U(N)$ and $U(k)$ matrices restricted to
the invariant blocks (\ref{inv}).

The results for O3$^+$ can be read off from
(\ref{inv}) by exchanging symmetric and antisymmetric
representations.

\section{ADS-like superpotential}

\subsection{D3-D(-1) one-loop vacuum amplitudes }
\label{sd3d1oneloop}

Non-perturbative superpotentials can be computed from the
instanton moduli space integral \cite{Dorey:2002ik,Billo:2002hm,Blumenhagen:2006xt}
\beq
S_W=   e^{  \langle \one
\rangle_{\cD} +  \langle \one \rangle_{\cA}+   \langle \one
\rangle_{\cM} }
\,  \mu^{\beta_n k_n }\, \int_{\mathfrak{M}} e^{-S_{k,N}-S_\vf}
  \label{intads0}
\eeq
The integration is over the instanton moduli space, ${\mathfrak{M}}$,
$\langle \one \rangle_\cD$ is the disk amplitude and
 $\langle \one \rangle_{\cA, \cM}$ are the one-loop
 vacuum amplitudes with at least one end on the D(-1)-instanton.
 The factor  $\mu^{\beta_n k_n }$, $\mu$ being the energy scale, comes from the
quadratic fluctuations around the instanton background and as we will see
it combines with a similar contribution coming from the moduli measure to give a dimensionless $S_W$.

The terms in front of the integral in (\ref{intads0}) combine into
\be
S_W=    \Lambda^{\beta_n k_n }\, \int_{\mathfrak{M}} e^{-S_{k,N}-S_\vf}
\label{ebeta}
\ee
with
\be
 \Lambda^{k_n \beta_n}=e^{2\pi i k_n \tau_n(\mu)} \,  \mu^{\beta_n k_n }\quad\quad
\tau_n(\mu)=\tau_n-{\beta_n\over 2\pi i }\,\ln\frac{\mu}{\mu_0}
\ee
the one-loop renormalization group invariant and the running coupling constant respectively
and $\tau_n$ refers to the complexified coupling constant of the $n^{\rm th}$ gauge group.
$\mu_0$  is a reference scale.

More precisely, the disk amplitude and one-loop amplitudes yields
\bea
 e^{ \langle \one
\rangle_{\cD} } &=& e^{2\pi i k_n \tau_n} \quad\quad \tau_n={\theta_n \over 2\pi}+{4\pi i \over
g_n^2}
\nn\\
e^{     \langle \one \rangle_{\cA}+   \langle \one
\rangle_{\cM} } &=& \left({\mu\over \mu_0}\right)^{-\beta_n \k_n}+\ldots\label{runningb}
\eea
with dots refering to threshold corrections that will not be considered here.

To see (\ref{runningb}) we should compute the following one-loop amplitudes
     \bea
  \langle \one \rangle_{\cA} &=& -\int_{\mu_0}^\mu {dt\over t} \,{1\over 12} {\rm Tr}[
  (1+(-)^F)(1+\theta+\theta^2)\, q^{L_0-a}] \nn\\
&=& - \int_{\mu_0}^\mu {dt\over t} \cA_{D(-1)D3}=- \cA_{0,D(-1)D3} \ln {\mu\over \mu_0}+\ldots
    \nn\\
   \langle \one \rangle_{\cM} &=&- \int_{\mu_0}^\mu {dt\over t} {1\over 12} {\rm Tr}  [\,\Omega I\,
  (1+(-)^F)(1+\theta+\theta^2)\, q^{L_0-a}]\nn\\
&=&  -\int_{\mu_0}^\mu {dt\over t} \cM_{D(-1)}= -\cM_{0,D(-1)} \ln {\mu\over \mu_0}+\ldots
   \label{1am0}
  \eea
In the above formula $\mu$ enters as a UV regulator in the open string channel
 (see \cite{Bianchi:2000vb} for details)
and $\cA_0,\cM_0$ are the massless contributions to the amplitudes.

   We start by considering the O3$^-$ projection.
   It is important to notice that only the annulus with one end on the D(-1) and one on the D3
  contributes to these amplitudes since D(-1)-D(-1) amplitudes cancel due to the Riemann identity.
  One finds
       \bea
   \cA_{D(-1),D3} &=& {4\over 12} {\rm tr} \gamma_{\theta,k}{\rm tr} \gamma_{\theta,N}
   \sum_{\alpha,\beta}\, c_{\alpha \beta} {\eta^3\over
   \vartheta [^\alpha_\beta]}  { \vartheta [^{\alpha+{1\over 2}}_\beta]^2\over
   \vartheta [^{0}_{1\over 2}]^2} \prod_{i=1}^3 { \vartheta [^\alpha_{\beta+h_i}]\over
   \hat{\vartheta} [^{1\over 2}_{{1\over 2}+h_i}]}  \nn\\
&=&     \ft32 (k_0-k_1)(N_0-N_1) +\ldots\nn\\
     \cM_{D(-1)}&=&   {2\over 12} {\rm tr} \gamma_{\theta^2,k}
   \sum_{\alpha,\beta}\, c_{\alpha \beta} {\eta^3\over
   \vartheta [^\alpha_\beta]}  { \vartheta [^\alpha_{\beta+{1\over 2}}]^2\over
  \hat{ \vartheta} [^{1\over 2}_{0}]^2} \prod_{i=1}^3 { \vartheta [^\alpha_{\beta+h_i}]\over
  \hat{ \vartheta} [^{1\over 2}_{{1\over 2}+h_i}]}  \nn\\
&=&     -3 (k_0-k_1)  +\ldots\label{1am}  \eea
 The sum runs over the even spin structures and $c_{\alpha \beta}=(-)^{2(\alpha+\beta)}$.
 The term ${\eta^3\over
   \vartheta [^\alpha_\beta]} $ comes from the $(b,c)$ and $(\beta,\gamma)$
   ghosts while the extra five thetas in the
 numerator and denominator describe the contributions of the ten fermionic and bosonic
 worldsheet degrees of freedom. We adopt the shorthand notation
 $\hat{\vartheta}[^{1\over 2}_h]\equiv \vartheta[^{1\over 2}_h]/(2\cos\pi h)$ to
 describe the massive contribution of a periodic boson to the partition function.
 $h_i=(\ft13,\ft13,-\ft23)$ denote the $\Z_3$-twists
 while the extra $\ft12$-shifts in the annulus  account
 for the D(-1)-D3 open string twist along Neuman-Dirichlet directions
 while $\ft12$ twists in the M\"obius come from the $I$-projection.
In addition we used the fact that the contribution of the unprojected sector is zero
 after using the Riemann identity while that of the $ \theta$- and $\theta^2$-projected sectors
  are identical explaining the overall factor of 2. The extra factor of 2 in the annulus
  comes from the two orientations of the string.
   The second line displays the massless contributions.  We use the Chan Paton
   traces
 \bea
    {\rm tr} \gamma_{\one,k}&=&k_0+2k_1\quad\quad {\rm tr} \gamma_{\theta,k}=k_0-k_1
    \nn\\
     {\rm tr} \gamma_{\one,N}&=&N_0+2N_1\quad\quad {\rm tr} \gamma_{\theta,N}=N_0-N_1
\eea
 that follows from (\ref{gz3})
and the first few terms in the theta expansions
  \bea
       \vartheta [^0_h]&=&1+q^{1\over 2}\, 2\cos 2\pi h+\ldots \quad \quad
         \vartheta [^{1\over 2}_h]=q^{1\over 8}\, 2\cos \pi h+\ldots\nn\\
          \eta &=& q^{1\over 24}+\ldots
  \eea
   From (\ref{1am}) one finds
   \be
  {\cal A}_0+{\cal M}_0=\ft32(k_0-k_1)(N_0-N_1-2)    =k_n \beta_n \label{a0m01}
    \ee
  with
$\beta_n$ the one-loop $\beta$ coefficients given in (\ref{betaG}).
Plugging (\ref{a0m01}) into (\ref{1am0})  results into (\ref{runningb}).
The fact that the $\beta$ function coefficients are reproduced by the instanton vacuum amplitudes is a
  nice test of the instanton field content (\ref{inv}).

 Now let us determine the dependence of the instanton measure  on the string scale $M_s\sim \alpha^{\prime\, -1/2}$.
The scaling of the various instanton moduli follows from (\ref{Sd}):
\beqn
&& D, g_0\sim M_s^{2} \quad\quad  \chi_a, \vf_a \sim M_s \quad\quad w_{\dalpha}, a_\mu \sim M_s^{-1} \nn\\
 && \nu^A,~ \theta^A_{\alpha}\sim M_s^{-{1/2}}\quad\quad
 \bar{\theta}_{A\dalpha}\sim M_s^{3/ 2}\label{scalings1}
\eeqn
Collecting from (\ref{inv}) the number of components of the various moduli entering in the
instanton measure one finds \footnote{We recall that fermionic differentials
scale as the inverse of the dimension of the fermion itself. This explains the extra minus sign in
(\ref{scaling}).}
  \beqn
  \int_{\mathfrak{M}} \, e^{-S_{k,N}-S_\vf} &\sim & M_s^{-\beta_n k_n} \nn\\
  k_n \beta_n  &=& -2 n_{D}-n_{\chi}+n_a+n_w+\ft32 n_{\bar{\theta}}-\ft12 n_{\theta}-\ft12 n_\nu\nn\\
   &=& \ft32 (k_0-k_1)(N_0-N_1-2)  \label{scaling}
\eeqn
Notice that this factor  precisely combines with that  in (\ref{ebeta}) leading to a dimensionless $S_W$
as expected. This simple dimensional analysis can be used to determine the form of the
allowed ADS superpotentials in the gauge theory.
A superpotential is generated if and only if the integral over the instanton moduli space
reduce to an integral over $x_0^\mu$
describing the center of the instanton and
$\theta_\alpha$ its superpartner.
More precisely
 \beq S_W=  \Lambda^{k_n \beta_n }\,
  \int_{\mathfrak{M}}
e^{-S_{k,N}-S_\vf} =
 c\, \int d^4x_0 d^2 \theta \, \frac{\Lambda^{k_n \beta_n} }{\vf^{k_n \beta_n -3}}
 \label{intads}
\eeq
where $c$ is a numerical constant.
Whether $c$ is zero or not depends
on the presence or not of extra fermionic zero modes besides $\theta$.
Notice that the power of $\varphi$ is completely fixed requiring that $S_W$ is
dimensionless.
 The precise form of the
superpotential requires the evaluation of the moduli space
integral and will be the subject of the next section. The
superpotential follows from (\ref{intads}) after promoting $\vf^I$
to the chiral superfield $\Phi^I$ and $x_0,\theta_\alpha$ to the
measure of the superspace
 \beq
  S_W=c\,
  \int d^4x d^2 \theta \,{\Lambda^{k_n\beta_n }\over \Phi^{k_n\beta_n -3}}
\label{ads}
 \eeq
  A superpotential of type (\ref{ads}) is generated whenever
  \cite{Veneziano:1982ah,Taylor:1982bp,Affleck:1983rr,Affleck:1983mk}
\beq
 \langle   \lambda^2 \, \varphi^{k_n\beta_n -3}\rangle \neq 0 \label{nonzero}
  %\quad\quad b_n k_n-3>0
\eeq
 %Thanks to the Yukawa coupling
%\beq L_{Yuk} = g_{YM}
%\varphi^\dagger \psi \lambda \label{yukcoup}
%\eeq
 Each scalar $\varphi$ soaks two fermionic zero-modes
and each gaugino $\lambda$ one zero mode\footnote{This can be seen
by explicitly solving the equations of motion of the gaugino and
the $\vf$-field in the instanton background  \cite{Dorey:1999pd}.
In particular the source for the scalar field comes from the
Yukawa coupling $L_{Yuk} = g_{YM} \varphi^\dagger \psi \lambda$ in
the gauge theory action. }. The condition (\ref{nonzero})
translates into
 \beq
{\rm dim} {\mathfrak{M}}_F=2k_n\beta_n -4
%=18(k_1-k_0)-4
\label{dimF}
 \eeq
 with ${\rm dim} {\mathfrak{M}}_F$ the fermionic dimension of the instanton super-moduli space.
 The number of fermionic zero modes can be read off from (\ref{inv})
 \beqn
 {\rm dim} {\mathfrak{M}}_F &=& n_\theta+n_\nu-n_{\bar{\theta}}\nn\\
 &=& k_0(3 N_1+N_0-2)+k_1[2 N_1+3 (N_0 + N_1-2)]\nn\\
 &=& k_0(4 N_0+10)+k_1(8 N_0+14)\label{dimF2}
\eeqn
 where we used the fact that $\bar{\theta}_{\dalpha A}$ enter as a Lagrangian multiplier
imposing the fermionic ADHM constraint and therefore subtracts degrees of freedom.
 The last line in (\ref{dimF2}) follows from using the anomaly cancellation
 condition $N_1=N_0+4$.
The result (\ref{dimF2}) is consistent with the Atiyah-Singer index theorem that states
\beqn
 {\rm dim} {\mathfrak{M}}_F &=&  2k_0 \left[\ell({\bf \ft12 N_0(N_0-1)})+3 N_1 \ell({\bf N_0})\right]\nn\\
 &&+  2k_1 \left[\ell({\bf N_1 \bar{N_1}})+3 N_0 \ell({\bf N_1})+3  \ell({\bf \ft12N_1(N_1-1)})\right]\nn\\
 &=& k_0(3 N_1+N_0-2)+k_1[2 N_1+3 (N_0 + N_1-2)]
 \eeqn
Combining (\ref{dimF}) and (\ref{dimF2}) one finds
 \beq
N_0={k_1-7 k_0-1\over k_0+2 k_1}
 \eeq
 One can easily see that the only non-negative solution for $N_0$ is
 $$
 N_0=0 \quad\quad k_0=0 \quad\quad  k_1=1
 $$
We conclude that in the class of $U(N_0+4)\times SO(N_0)$ SYM
theories describing the low-energy dynamics of D3-branes
on the $\Z_3$ orientifold only the $U(4)$ theory with three
chiral multiplets in the antisymmetric leads to an ADS-like
superpotential generated by gauge instantons.

The counting can be easily repeated for the $Sp(N_1+4)\times
U(N_1)$ cases by exchanging symmetric and antisymmetric
representations in (\ref{inv}) as required by the presence
of the O3$^+$-plane. The results are \beqn
 k_n \beta_n &=& 9(k_0-k_1) \nn\\
 {\rm dim} {\mathfrak{M}}_F &=& k_0(4 N_1+6)+k_1(8 N_1+18) \nn\\
N_1 &=& {3 k_0-9 k_1-1\over k_0+2 k_1}
 \eeqn
 One can easily see that the only non-negative solution is
 $$
 N_1=2 \quad\quad k_0=1 \quad\quad k_1=0
 $$
We conclude that in this class, only the gauge theory $Sp(6)\times
U(2)$ with three chiral multiplets in the $(\tinyyoung{\hfill},\bar{\tinyyoung{\hfill}})
+(\bullet,\tinyyoung{\hfill,\hfill})$ admits an ADS-like superpotential
generated by instantons.

The aim of the rest of this section is to compute $S_W$.
The integral (\ref{intads}) will be evaluated in turn for the $Sp(6)\times U(2)$ and $U(4)$ case.

\subsection{$Sp(6)\times U(2)$ superpotential }

 We first consider the O3$^{+}$ case, \ie the $Sp(6)\times U(2)$ gauge theory
with three  chiral multiplets in the $\left[({\bf 6},{\bf
\bar{2}})+({\bf 1},{\bf 3})\right]$.
 The instanton moduli is given by (\ref{inv}) after flipping symmetric/antisymmetric representations
 in order to deal with the symplectic  projection. Plugging
 $k_0=1,k_1=0,N_0=6,N_1=2$  into (\ref{inv}) one finds the the surviving fields
 \beqn
   ~a_\mu,~ w_{u_0}^{\dalpha}, \theta_\alpha^0,~ \nu^{0 u_0}, ~\nu^{I u_1}
 \eeqn
 with $u_0=1,..6$, and $u_1=1,2$, whose position from lower to upper has been switched
in this section for notational convenience as we will momentarily
see.
 In particular both
 $\bar{\theta}_{0\dalpha}$ and $D^c$ are projected out
 (the D(-1) ``gauge'' group is $O(1)\approx \Z_2$ in this case)
 and therefore no ADHM constraint survives.
 The instanton action reduces then to
\beq
 S=S_K+S_\vf=w^{u_0}_{\dalpha} \bar{\vf}_{I u_0 u_1 } \vf^{I\,u_1 v_0}  w_{v_0}^{ \dalpha}+
 \nu^{I u_1} \nu^{0 u_0} \bar{\vf}_{I u_0 u_1}
\eeq
 Here and below we omit numerical coefficients that can be always reabsorbed at the end
 in the definition of the scale.
  The integrations over $w_{u_0}^{\dalpha}$, $\nu^0_{u_0},\nu^I_{u_1}$ are gaussian and
the final result, up to a non vanishing numerical constant, can be
written as
\beqn
S_W &=& \Lambda^{9}\,\int d^4
a d^2\theta \, {{\rm det}_{6\times 6}
\,(\bar{\vf}_{I u_1 , u_0})\over {\rm det}_{6\times 6} \,  (\bar{\vf}_{I u_1 , u_0} \vf^{I u_1 , v_0}) }\nn\\
&=& \int d^4 a d^2\theta \,
{\Lambda^{9} \over {\rm det}_{6\times 6} \,(\vf^{I u_1 , u_0})  }
\label{potsp6}
\eeqn
where
we have exploited the possibility of combining $I$ and $u_1$ in
one `bi-index' $I u_1$ so as to get a range of six values.
For the sake of simplicity we have
 dropped the subscript $0$ denoting bare scalar fields.
In the following scalar fields entering in formulae involving $\Lambda$ will be
always understood to be bare.
The last step makes use of ${\rm det}(A B) = {\rm det}(A) {\rm
det}(B)$.

\subsection{ $U(4)$ superpotential }

We now consider the O3$^{-}$ case, \ie the $U(4)$ gauge theory
 with three chiral multiplets in the ${\bf 6}$.
Setting $k_0=0,k_1=1,N_0=0,N_1=4$ in (\ref{inv})
 the surviving fields can be written as
 \beqn
&&  \vf^{I[u v]}_{(0)} \: , \: \bar\vf_{I[u v](0)}\: , \:
a_{\mu(0)} \: , \: \bar\chi_{I(-2)} \: , \: {\chi}^I_{(+2)}\: , \:
D^c_{(0)} \: , \:
w_{u(+1)}^{\dalpha}\: , \: \bar{w}_{\dalpha(-1)}^{u} \nn\\
  && \theta^0_{\alpha (0)}
 \: , \:  \bar{\theta}_{0\dalpha(0)}\: , \:  \bar{\theta}_{\dalpha I(-1)}
 \: ; \:   \nu^{0}_{u(+1)}\: , \:  \bar{\nu}^{0 u}_{(-1)} \: , \:  \nu^{I
 u}_{(+1)}
  \label{inv 2 }
 \eeqn
 with $u=1,..4$ and the charge $q$ under $U(1)_{k_1}$ is denoted in parentheses.
Plugging into (\ref{Sd}) (after taking $\alpha'\to 0$) one finds
\be S = S_B + S_F \ee where \bea
 S_F &=& \left(\bar{\nu}^{0 u} w_{u \dot{\alpha}}+
\bar{w}^u_{\dalpha} \nu^0_u \right)
\bar{\theta}_0^{\dot{\alpha}}+{\nu}^{Iu}\, w_{u\dot{\alpha}}\,
\bar{\theta}_I^{\dot{\alpha}} + \bar{\chi}_I \nu^0_u \nu^{Iu} +
{\nu}^{I u}\bar\vf_{I uv} \bar\nu^{0 v} \nn\\
 S_B  &=&
\bar{w}^u_{\dalpha}   \bar\vf_{I u w} \vf^{I w v} w_v^{\dalpha} +
 \vf^{I u v} w_u^{\dalpha} w_{v\dalpha}\bar\chi_I +
 \bar\vf_{I u v} \bar{w}^{u\dalpha} \bar{w}^v_{\dalpha}\chi^I +
 \bar{w}^{u\dalpha} w_{u\dalpha} \bar{\chi}_I
 \chi^I\nn\\
 &&+ D^{c}\,\bar{w}
\sigma_c w
\eea
 As before we omit numerical coefficients.
 The integral over $D^c$ leads to a $\delta$ function on the ADHM constraints
 \be
 \int d^{8} w d^8 \bar{w} \delta^3(\bar{w}
\sigma_c w )=\int d\rho \, \rho^9 d^{12} {\cal U} \label{duw}
 \ee
In the r.h.s of  (\ref{duw}) we have solved the ADHM constraints in favor of $w$ and ${\cal U}$
defined by
\be
w_{u\dalpha}=\rho\, {\cal U}_{u\dalpha} \quad\quad
\bar{w}^{u\dalpha}=\rho\, {\cal \bar{U}}^{u\dalpha} \quad\quad
 {\cal \bar{U}}^{u\dalpha} {\cal U}_{u\dbeta}=\delta^{\dalpha}_{\dbeta}
\ee
 The coset representatives ${\cal U}_{\dalpha u}$ parameterizes the $SU(4)/SU(2)$ orientations of
 the instanton inside the gauge group.
 The fermionic integrations lead to the determinant
 \be
 {\Delta}_{\rm F}=\rho^8 \, \epsilon^{u_1 u_2 u_3 u_4}
 \epsilon^{v_1 v_2 u_5 u_6}\epsilon^{v_3 v_4 v_5 v_6}
 X_{u_1 v_1 u_2 v_2}\, X_{u_3 v_3 u_4 v_4}\,
 Y_{u_5 v_5}\, Y_{u_6 v_6}\label{detf}
 \ee
with
\bea
X_{u_1 v_1 u_2 v_2} &=&\epsilon^{I_1 I_2 I_3} \bar{\chi}_{I_1} \bar{\vf}_{I_2 u_1 v_1} \bar{\vf}_{I_3 u_2 v_2}\nn\\
Y_{uv} &=&   {\cal U}^{\dalpha}_u {\cal U}_{\dalpha u} \eea The
bosonic integrals are more involved. For arbitrary choices of the
scalar VEV's $\bar\vf_{I}$ and $\vf^{I}$, even along the flat
directions of the potential, the integration over $\cU$ represents
a challenging if not a prohibitive task. Fortunately choosing
$\vf^{I u v} = \vf \eta^{I u v}$, $\bar{\vf}_{I u v} = \bar{\vf}
\eta^{I u v}$, the full $\vf$-dependence can be factorized.
 $SU(4)$ gauge and $SU(3)$ `flavor' invariance can then be used to
 recover the full answer.
  After the  rescaling
 \be
 \rho^2\to \rho^2/(\vf \bar{\vf})\quad\quad \chi^I\to \vf\chi^I\quad\quad   \bar{\chi}_I\to \bar{\vf}\bar{\chi}_I
 \ee
    The integral becomes
 \be
 S_W= \Lambda^{9}\, I\, \int d^4x_0 d^2 \theta\, {1\over \vf^6 }\,
\label{resads}
 \ee
with $I$ the $\varphi$-independent integral
  \bea
I &=&   \int  \, d\rho \rho^9\, d^{12}{\cal U}\, d^3\chi d^3\bar{\chi}\,
  {\Delta}_{\rm F} \, e^{-\tilde{S}_B}\nn\\
    \tilde{S}_B & =&
-\rho^2(1+
 \eta^{I u v}  Y_{uv}  \chi_I +
 \bar\eta_{I u v} \bar{Y}^{uv} \chi^I +
 \bar{\chi}_I
 \chi^I)\eea
 and ${\Delta}_{\rm F} $ given again by  (\ref{detf}) but now in terms of
 \be
 X_{u_1 v_1 u_2 v_2} =\epsilon^{I_1 I_2 I_3} \bar{\chi}_{I_1}
 \bar{\eta}_{I_2 u_1 v_1} \bar{\eta}_{I_3 u_2 v_2}\nn\\
  \ee
  Finally one can restore the gauge covariance of  (\ref{resads}) by noticing that there is a unique
  $SU(4)_c\times SU(3)_f$ singlet in the symmetric tensor of six $\vf^I$
  $$
  {\rm det}_{3\times 3} [ \epsilon_{u_1..u_4} \vf^{I u_1 u_2}
\vf^{J u_3 u_4} ]
  $$
  Therefore one can replace $\vf^6$
  in (\ref{resads}) by this singlet. The superpotential follows after replacing $\vf^I\to
  \Phi^I$
  \be
  S_W=
c \int d^4x d^2 \theta\, {\Lambda^{9} \over
{\rm det}_{3\times 3} [ \epsilon_{u_1..u_4} \Phi^{I u_1 u_2}
\Phi^{J u_3 u_4} ]}
\ee
where $c$ is a computable non-zero
numerical coefficient.

\section{ED3-instantons}
 \label{newinst}

Let us now consider the ED3-D3 system. We restrict ourselves to
the compact case $T^6/\Z_3$ and consider the ED3 fractional
instanton wrapping a four-cycle ${\cal C}_n$ inside $T^6/\Z_3$.
  We start by considering the O3$^-$-orientifold projection. The zero
modes of the Yang-Mills fields in the instanton background can be
described as before in terms of open strings with at least one end
on the ED3.
 Open strings connecting ED3 and D3 branes have 8 Neumann-Dirichlet
 directions therefore the zero-mode dynamics of the ED3-D3 system
 is equivalent to that of the D7-D(-1) bound state.
 The instanton action can be found starting from that of the $\cN = (8,0)$
sigma model describing the low energy dynamics of a D1-D9 bound state
  in type I \cite{Gava:1998sv} reduced down to zero dimensions.
  In flat space the D(-1)-D7 action reads
 \be
 S = {\rm tr}_k \left[{1\over g_0^2}\,S_g+ S_K+S_D \right] \label{SEd}
\ee
with
\bea
 S_g &=&- [\chi,\bar{\chi}]^2+\tilde{\Theta}^{\dot a} \chi \tilde{\Theta}^{\dot a}+D^c D^c\nn\\
 S_K &=& -[\chi,X_m][\bar{\chi},X_m]+ \Theta^a \bar{\chi} \Theta^a+ \nu (\chi+\vf)
 \nu\nn\\
S_D &=& \tilde{\Theta}^{\dot a} X_m \Gamma_{\dot{a} a}^m \Theta^a+
D^c \hat{\Gamma}^c_{mn} [X_m,X_n] \label{SdED}
 \eea
 with $m=1,\ldots,8_v$, $a=1,\ldots,8_s$, $\dot{a}=1,\ldots,8_c$, $c=1,,\ldots,7$.
We denote by $\varphi = m_I (\cC_n) \varphi^I$, the gauge scalar parametrizing the
position of the D3-brane along the direction perpendicular to the 4-cycle $\cC_n$.
Here $\Gamma_{\dot{a} a}^m$,
  $\hat{\Gamma}^c_{mn}$ are gamma matrices of $SO(8)$ and $SO(7)$ respectively.
 The introduction of the auxiliary fields $D^c$ has broken the
 manifest $SO(8)$ invariance of the  action that will be further broken by the
 $\Z_3$-projection.
In (\ref{SdED}), $X_m$ and $\chi,\bar{\chi}$  describe the
position of the D(-1)-instanton in the directions longitudinal and
perpendicular to the D7-brane respectively while
$\Theta^a,\tilde{\Theta}^{\dot a}$ are the fermionic superpartners
grouped according to the their chirality along the
Dirichlet-Dirichlet $\chi$-plane. Unlike the D(-1)-D3 case, in the
case of  8 Neumann-Dirichlet directions $\Omega$ acts in the same
way on the D(-1) and D7 Chan-Paton indices. This implies that
$D^c$ transform in the adjoint of $SO(k)$ if we take the D7 gauge
symmetry to be $SO(N)$. In addition $I$ acts as \be I: \quad\quad
X_m \to -X_m \quad\quad \Theta^a \to -\Theta^a \label{IED} \ee
   Fields with eigenvalues $\Omega I=-$ are then in the following
representations of $SO(k)\times   SO(N)$
 \bea
( \chi,\bar{\chi}, D^c ,\tilde{\Theta}^{\dot{a}} ) &&   {\bf \ft12k(k-1)}\nn\\
( X_m , \Theta^a)  &&   {\bf \ft12k(k+1)}\nn\\
\nu  &&  {\bf k}{\bf N}\label{ed3inv0}
 \eea
  Fields even under $I$ transform in the adjoint of $SO(k)$ while odd fields
tranform in the symmetric representation.
  For $k=1$, $N=32$ the D(-1)-D7 system or equivalently the D1-D9 bound state
  describes the S-dual version of the fundamental heterotic string on $T^2$.
  $k>1$ bound states correspond to multiple windings of the heterotic string
   \cite{Gava:1998sv}.

The field $D^c$ implements the one-real $D$ and three complex $F$
flatness conditions
\be
V=-{1\over g_0^2}\sum_{c=1}^7 D^c D^c = -g_0^2 \sum_{m,n=1}^8
[X_m,X_n]^2=0
\ee
with
\be D^c =-\ft12 g_0^2 \Gamma^c_{mn} [X_m,X_n] \ee An
explicit choice of $\Gamma$ matrices in $D=7$ is given by
($a=1,2,3$) \beq \Gamma^a_{8\times 8} = i \sigma_1 \otimes
\eta^a_{4\times 4} \quad \Gamma^{a+3}_{8\times 8} = i \sigma_3
\otimes \bar\eta^a_{4\times 4} \quad \Gamma^7_{8\times 8} = i
\sigma_2 \otimes {\bf 1}_{4\times 4} \label{gamma7}
\eeq
  As in section \ref{sd1instanton}
$\Z_3$ acts both on spacetime and Chan-Paton indices.  Chan-Paton indices decompose
  as $N\to N_0+N_1+\bar{N_1}$ and $k\to k_0+k_1+\bar{k}_1$.
    Spacetime indices on the other hand decompose as
 \bea
 8_v &=& 4+2_\omega+2_{\bar{\omega}}\nn\\
  8_s &=& 2+2_\omega+4_{\bar{\omega}}\nn\\
  8_c &=& 2+2_{\bar{\omega}}+4_\omega\nn\\
   7 &=& 3+2_\omega+2_{\bar{\omega}}
 \eea
  In addition $\chi,\nu$ transform with eigenvalue $\omega$ under $\Z_3$.  Combining with
   (\ref{ed3inv0}) one finds the $\Z_3$-invariant components
\bea
 \chi,\bar{\chi}  &&  {\bf \ft12k_1(k_1-1)}+{\bf k_0 \bar{k}_1}+{\rm h.c.}  \nn\\
D^c  && 3({\bf \ft12k_0(k_0-1)}+{\bf k_1 \bar{k}_1} ) +2 \left[
{\bf \ft12k_1(k_1-1)}+ {\bf k_0 \bar{k}_1}
+{\rm h.c.}\right]\nn\\
 \tilde{\Theta}^{\dot{a}}  &&    2\left[{\bf \ft12k_0(k_0-1)}+{\bf k_1 \bar{k}_1}\right]
 +2\left[{\bf \ft12\bar{k}_1(\bar{k}_1-1)}+{\bf k_0 k_1})\right]\nn\\
&& +4 \left[({\bf \ft12 k_1(k_1-1)}+{\bf k_0 \bar{k}_1})\right]\nn\\
 X_m  && 4\left[ {\bf \ft12k_0(k_0+1)}+ {\bf k_1\bar{k}_1}\right]
 +2 \left[{\bf \ft12k_1(k_1+1)}+{\bf k_0 \bar{k}_1}
+{\rm h.c.}\right]\nn\\
 \Theta^a  &&  2\left[{\bf \ft12k_0(k_0+1)}+{\bf k_1 \bar{k}_1}\right]
 +2\left[{\bf \ft12k_1(k_1+1)}+{\bf k_0 \bar{k}_1})\right]\nn\\
&& +4 \left[({\bf \ft12 \bar{k}_1(\bar{k}_1+1)}+{\bf k_0 k_1})\right]   \nn\\
\nu && {\bf k_0 \bar{N}_1}+{\bf k_1 N_1}+{\bf \bar{k}_1 N_0}
\label{ed3inv}
 \eea

\subsection{D3-ED3 one-loop vacuum amplitudes }

ED3 generated superpotentials can be computed  following the same
steps as in section \ref{sd3d1oneloop}. The
disk amplitude can be written as
\be e^{  \langle \one \rangle_{\cD}
}=e^{2\pi i k_n \tilde{\tau}_n} \quad\quad \tilde{\tau}_n=i\,{4\pi
 V_4(\cC_n) \over g_n^2 \,\alpha^{\prime\, -2} }+\int_{\cC_n} (C_4+C_0\wedge
R\wedge R)
\ee
 $\tilde{\tau}_n$ describes the coupling of closed string moduli to the ED3
instanton wrapping the 4-cycle $\cC_n$ with volume $ V_4(\cC_n)$.
We remark that  closed string states in the $\Z_3$-twisted
sectors flow in the ED3-ED3 cylinder amplitude and therefore
$\tilde{\tau}_n$ is function of both untwisted and twisted closed twisted moduli.
This is not surprising since the volume of the cycle depends also on the volume
of the exceptional cycles that the ED3 wraps.

 The annulus and M\"obius amplitudes are given by
      \bea
 &&  \cA_{ED3,D3} = {2\over 12}
   \sum_{\alpha,\beta}\, c_{\alpha \beta} {\eta^3\over
   \vartheta [^\alpha_\beta]}  { \vartheta [^{\alpha+{1\over 2}}_\beta]^2\over
   \vartheta [^{0}_{1\over 2}]^2} \times
   \nn\\
&&   \times \left( 2 {\rm tr} \gamma_{\theta,k}{\rm tr} \gamma_{\theta,N}
     { \vartheta [^{\alpha+{1\over 2}}_{\beta+h_1}]^2\over
   \vartheta [^{0}_{{1\over 2}+h_1}]^2}
    { \vartheta [^\alpha_{\beta-2h_1}]\over
  \hat{ \vartheta} [^{1\over 2}_{{1\over 2}-2h_1}]}
    +     {\rm tr} \gamma_{\one,k}{\rm tr} \gamma_{\one,N}
      { \vartheta [^{\alpha+{1\over 2}}_{\beta}]^2\over
   \vartheta [^{0}_{{1\over 2}}]^2}
     { \vartheta [^\alpha_{\beta}]\over
   \eta^3} \right) \nn\\
&& =     -\ft12 k_0 N_1-\ft12 k_1(N_0+N_1) +\ldots\nn\\
   &&  \cM_{ED3}= -{1\over 12}
   \sum_{\alpha,\beta}\, c_{\alpha \beta} {\eta^3\over
   \vartheta [^\alpha_\beta]}  { \vartheta [^{\alpha}_{\beta+{1\over 2}}]^2\over
   \hat{\vartheta} [^{1\over 2}_{0}]^2} \times
   \nn\\
&&   \times \left( 2 {\rm tr} \gamma_{\theta^2,k}
     { \vartheta [^{\alpha }_{\beta+{1\over 2}+h_1}]^2\over
   \hat{\vartheta} [^{1\over 2}_{0+h_1}]^2}
    { \vartheta [^\alpha_{\beta-2h_1}]\over
   \hat{\vartheta} [^{1\over 2}_{{1\over 2}-2h_1}]}
    +     {\rm tr} \gamma_{\one,k}
      { \vartheta [^{\alpha}_{\beta+{1\over 2}}]^2\over
   \hat{\vartheta} [^{1\over 2}_0]^2}
     { \vartheta [^\alpha_{\beta}]\over
   \eta^3} \right) \nn\\
&& =   3 k_0+ k_1  +\ldots \label{1amED}
\eea
 The origin of the various contributions is the same that in
the D(-1)-D3 system. Now the D3-ED3 open strings have 8
Neumann-Dirichlet directions explaining the extra $\ft12$ twists
in the annulus amplitude. On the other side, the  $I$ projection
accounts for the  $\ft12$-shift in the M\"obius amplitude.
Notice that unlike the D(-1)-D3 case, the unprojected amplitude
${\rm tr} \one$, now gives a non-trivial contribution.

 Collecting the contributions from (\ref{1amED}) one finds
   \be
  \tilde{\Lambda}^{k_n b_n}
  = \mu^{k_n b_n}\,e^{ \langle \one \rangle_{\cD}+ \langle \one \rangle_{\cA}+   \langle \one \rangle_{\cM} }
    =\mu^{k_n b_n}\,e^{2\pi i k_n\tilde{\tau}_n(\mu)}
    \label{ebetaED}
  \ee
with
\be
\tilde{\tau}_n(\mu)=\tilde{\tau}_n-{b_n\over 2\pi i}\,\ln\frac{\mu}{\mu_0}
\ee
and
  \be
  {\cal A}_0+{\cal M}_0=
k_n b_n  =\ft12 k_0(6-N_1)+\ft12 k_1(2-N_0-N_1)
  \ee
The interpretation of the $b_n$ as the one-loop $\beta$
function coefficients of the $\tilde{\tau}_n$ coupling, though
tantalizing, is not clear to us.
We will now check that $k_n b_n$ reproduces the right scale dependence of the instanton
measure.
The scaling of the various instanton moduli follows from (\ref{SdED}):
\beqn
&& D, g_0\sim M_s^{2} \quad\quad  \chi,\bar{\chi}, \vf \sim M_s \quad\quad X_m \sim M_s^{-1} \nn\\
 && \nu,~ \Theta^a \sim M_s^{-{1/2}}\quad\quad
 \tilde{\Theta}^{\dot a}\sim M_s^{3/ 2}\quad\quad
\eeqn

  Collecting from (\ref{ed3inv}) the number of degrees of freedom entering in the
   instanton supermoduli measure one finds
  \bea
  \int_{\mathfrak{M}} \, e^{-S_{k,N}} &\sim & M_s^{-k_n b_n } \nn\\
 k_n  b_n  &=& -2 n_{D}-n_{\chi}+n_X+\ft32 n_{\tilde{\Theta}}-\ft12 n_{\Theta}-\ft12 n_\nu\nn\\
   &=&\ft12 k_0(6-N_1)+\ft12 k_1(2-N_0-N_1) \label{scalingED}
\eea
  As in the previous case we write the instanton generated superpotential as the moduli space integral
\bea
S_W &=& \tilde{\Lambda}^{k_n b_n}\,
 \, \int_{\mathfrak{M}} e^{-S_{k,N}-S_\vf}=
 \int d^4x_0 d^2 \theta\,   \tilde{\Lambda}^{k_n b_n} \, \vf^{-k_n b_n+3}
  \label{intadsE}
\eea
 After promoting $\vf\to \Phi$ and $x_0,\theta_\alpha$ to the
measure of the superspace one finds the ED3 generated superpotential
 \beq
  S_W=
 \int d^4x d^2 \theta \,  \tilde{\Lambda}^{k_n b_n}\, \Phi^{-k_n b_n+3}
\label{adsE}
 \eeq
   The main difference with respect to the D(-1) instantons is that now
 $\varphi$ enters into $S_\vf$  (\ref{SdED}) only through the coupling to the
 $\nu$-fermions. This implies that in order to get a non zero result from the fermionic integral
in (\ref{intadsE}) only the $\nu$'s and the two fermionic zero modes
$\theta_\alpha \in \Theta^a$ should survive the orientifold
projections. {From} (\ref{ed3inv}) one can easily see that this implies
$k_0=1$, $k_1=0$. The same counting shows that no solutions are
allowed in the $Sp(N)$ case.

\subsection{The superpotential}

Here we evaluate the instanton moduli space integral for the
$SO(N_0)\times U(N_1)$ case. {From} our analysis above the
relevant cases are $k_0=1$, $k_1=0$.

The surviving fields in (\ref{ed3inv}) are \be \theta_\alpha \in
\Theta^a \quad\quad x_{0}^\mu \in X_m \quad\quad \nu_{u} \ee
 with $u=1,...N_1$. The instanton action reduces to
\be S = \nu_{u} \vf^{u v} \nu_{v}\label{sfed} \ee
 The superpotential is then given by the integral
 \bea
 S_W= \tilde{\Lambda}^{-{N_1\over 2}+3}
 \, \int d^4 x\, d^2\theta\, d^{N_1}\nu\, e^{-\nu \vf \nu}
\eea
 After integration over $\nu$ and lifting $\vf \to \Phi$ to the superfield
 one finds
 \be
S_W =  c\, \tilde{\Lambda}^{-{N_1\over 2}+3} \,
\int d^4 x\, d^2\theta\, \epsilon_{u_1....u_{N_1}} \Phi^{u_1
u_2}\Phi^{u_3 u_4}...\Phi^{u_{N_1-1} u_{N_1}} \label{swed3}
 \ee
  where $c$ is a non vanishing numerical constant. Notice that the result is non-trivial only when $N_1$ is even.
  The superpotentials (\ref{swed3}) are  non-renormalizable for $N_1>6$
 and grow for large vacuum expectation values where the low energy approximation
 breaks down. The only exceptions are
 \bea
   {\rm Majorana~masses} && U(4) +3\, \tinyyoung{\hfil,\hfil} \nn\\
 {\rm Yukawa~couplings}
 &&SO(2)\times U(6)+3\,  (\tinyyoung{\hfil},\bar{\tinyyoung{\hfil}})+3\,(\bullet,\tinyyoung{\hfil,\hfil})
  \eea
   Notice that both instanton generated Yukawa couplings  involve only the matter
   in the antisymmetric representation.

\section{ADS superpotentials: a general analysis}
\label{sads}
Here we consider a general
${\cal N}=1$ gauge theory with gauge group $U(N)$ and $n_{\rm Adj}$, $n_f/\bar{n}_f$,
$n_S/\bar{n}_S$, $n_A/\bar{n}_A$ number of chiral multiplets in the
adjoint, fundamental, symmetric and anti-symmetric
representations (and their complex conjugates) respectively.

  The cubic chiral anomaly, one-loop $\beta$ function and number of fermionic zero modes
  in the instanton background of the gauge theory can be written as
 \bea
 I_{\rm anom} &=& n_{f-}+n_{S-}(N+4)+n_{A-}(N-4)=0\label{ibdim}\\
\beta_{\rm 1-loop}&=& 3 N-N n_{\rm Adj}-\ft12 n_{f+}-\ft12 n_{S+}(N+2)-\ft12 n_{A+}(N-2)\nn\\
{\rm dim}{\mathfrak{M}}_F &=& k\left[2 N+ 2 N n_{\rm Adj}+
n_{f+}+n_{S+}(N+2) + n_{A+}(N-2)\right]\nn \eea
 with
 \be
 n_{f\pm}=n_f\pm \bar{n}_f \quad\quad n_{S\pm}=n_S\pm \bar{n}_S \quad\quad
 n_{A\pm}=n_A\pm \bar{n}_A \quad\quad
 \ee
 The condition for an Affleck, Dine and Seiberg like superpotential \cite{Affleck:1983rr,Affleck:1983mk}
 to be generated was determined in section \ref{sd3d1oneloop} to be
 \be
 {\rm dim}{\mathfrak{M}}_F=2 k \beta-4\label{conds}
 \ee
 Combining (\ref{ibdim}) and (\ref{conds}) one finds
 \bea
 \beta_{\rm 1-loop}&=& 2N+{1\over k}  \label{conds2} \\
 n_{f-} &=& -n_{S-}(N+4)- n_{A-}(N-4)\nn\\
 n_{f+} &=& 2 N-{2\over k}-2 N n_{\rm Adj}- n_{S+}(N+2)- n_{A+}(N-2)\nn
 \eea
   Remarkably the $\beta$ function in a theory admitting an instanton generated
   superpotential depends only on the rank of the gauge group.
  A simple inspection shows that a superpotential is generated only
  for $k=1$ and $n_{\rm Adj} =0$.  The complete list follows
  from a scan of any choice of $n_{S\pm}$,$n_{A\pm}$
  such that $n_+\geq |n_-|$ and $n_+ \geq 0$.
One finds
\bea && U(N)+ N_f\left(\, \tinyyoung{\hfil}+\bar{
\tinyyoung{\hfil} }\,\right)\quad\quad
~~~~~~N_f\leq N-1\nn\\
&& U(N)+
\tinyyoung{\hfil,\hfil}+
(N-4)\bar{ \tinyyoung{\hfil} }
+ N_f\left(\, \tinyyoung{\hfil}+\bar{ \tinyyoung{\hfil} }\,\right)\quad\quad N_f\leq 2\nn\\
&& U(4)+  2\,\tinyyoung{\hfil,\hfil}
+N_f \left(\, \tinyyoung{\hfil}+\bar{ \tinyyoung{\hfil} }\,\right)
\quad\quad ~~~~~N_f\leq 1\nn\\
&& U(4)+  3\,\tinyyoung{\hfil,\hfil} \nn\\
&& U(5)+  2\,\tinyyoung{\hfil,\hfil}+2 \bar{ \tinyyoung{\hfil} }
\eea
 The inequalities are saturated for gauge theories satisfying (\ref{conds}) and (\ref{conds2}),
while the lower cases are found by decoupling quark-antiquark
pairs via mass deformations.

  The generalization to $SO(N)/Sp(N)$ gauge groups is straightforward.
In these cases there is no restriction coming from anomalies since
representations are real. The $\beta$ function and the number of
fermionic zero modes in the instanton background are given by
 \bea
\beta_{\rm 1-loop}&=& \ft32(N\pm 2)-\ft12 n_{f}-\ft12 n_{S}(N+2)-\ft12 n_{A}(N-2)\nn\\
{\rm dim}{\cal M}_F &=& k\left[N\pm 2 +n_{f}+ n_{S}(N+2) +
n_{A}(N-2)\right]\nn \eea with upper sign for $Sp(N)$ and lower
sign for $SO(N)$ gauge groups.
 Imposing (\ref{conds}) one finds
\bea
 \beta_{\rm 1-loop}&=& N\pm 2+{1\over k}  \quad\quad    \\
 n_{f} &=& N\pm 2-{2\over k} - n_{S}(N+2)- n_{A}(N-2)\nn
 \eea
   The list of solutions is even shorter
 \bea
&& SO(N)+ N_f  \tinyyoung{\hfil} \quad\quad N_f\leq N-3\quad\quad k=2  \nn\\
&& Sp(N)+ N_f   \tinyyoung{\hfil}\quad\quad N_f\leq N\quad\quad k=1\nn\\
&& Sp(N)+ \tinyyoung{\hfil,\hfil}+2\,\tinyyoung{\hfil} \quad\quad k=1 \nn\\
 \eea
 Notice that $k=1$, respectively $k=2$, are the basic instantons in $Sp(N)$, respectively
 $SO(N)$,
since the instanton symmetry groups are in these cases $SO(k)$,
respectively $Sp(k)$.

\section{ Conclusions }
\label{conclusions}

In the present paper, we have given a detailed microscopic
derivation of non-perturbative superpotentials for chiral ${\cal
N}=1$  D3-brane gauge theories living at $\Z_3$-orientifold
singularities. We considered both unoriented projections leading
to $SO(N_1-4)\times U(N_1)$ and $Sp(N_1+4)\times U(N_1)$ gauge
theories with three generations of chiral matter in the
representations
 $(\tinyyoung{\hfil},\bar{\tinyyoung{\hfil}})+(\bullet,\tinyyoung{\hfil,\hfil})$
 and  $(\tinyyoung{\hfil},\bar{\tinyyoung{\hfil}})+(\bullet,\tinyyoung{\hfil\hfil})$
 respectively.

 The   $U(4)$  case was studied in details in  \cite{Bianchi:2007fx} and
 describes the local physics of type I theory
near the origin of $T^6/\Z_3$ with $SO(8)\times U(12)$ gauge group broken by
Wilson lines.
 In the present T-dual setting, there are two sources of non-perturbative effects: D(-1) and ED3 instantons.
 The former
realize the standard gauge instantons and lead to Affleck, Dine
and Seiberg
  like superpotentials. The latter lead to Majorana
  masses or non-renormalizable superpotentials  and were ignored till
very recently
\cite{Blumenhagen:2006xt,Haack:2006cy,Ibanez:2006da,Florea:2006si,Akerblom:2006hx,Bianchi:2007fx,Cvetic:2007ku,Argurio:2007}.

 Our explicit instanton computations confirm the form of ADS and stringy superpotentials
 proposed  in \cite{Bianchi:2007fx} on the basis of holomorphicity, dimensional analysis
  $U(1)$ anomaly and flavour symmetry.
  We show that ADS superpotentials are generated only
for the  $U(4)$ and $Sp(6)\times U(2)$  gauge theories in the $\Z_3$-orientifold list.
 The precise form  of the superpotential is derived from an integration over the
 instanton super-moduli space. Like in  \cite{Akerblom:2006hx},  the $\beta$ function running
 of gauge couplings are reproduced from  vacuum amplitudes given in terms of annulus and
 M\"obius
 amplitudes  ending on the instantons.
  The same analysis is performed for ``stringy instantons''  generated by Euclidean
ED3-branes (dual to ED1-strings in type I theory) wrapping
holomorphic four-cycles on $T^6/\Z_3$. A detailed microscopic
analysis of the multi-instanton super-moduli space encompasses
massless open string states with a least one end on the
ED3-instanton.
 We  show the generation of Majorana mass terms for the open string chiral
multiplets in the $U(4)$ case, Yukwa couplings for the $SO(2)\times U(6)$ gauge theory
  and  non-renormalizable
superpotentials for $SO(N_0)\times U(N_0+4)$ gauge theories.
 The field theory interpretation of the $\beta$ function coefficients generated by the
 one-loop vacuum amplitudes for open strings ending on the ED3-instantons is
 one of the most interesting open question left by our instanton super-moduli space analysis.
  As previously observed, the invariance under anomalous $U(1)$'s  results from a detailed
balance between the charges of the open strings involved and the
axionic shift of a closed string R-R modulus from the twisted
sector.

    Our present analysis has some analogies with the recent ones
\cite{Blumenhagen:2006xt,Haack:2006cy,Ibanez:2006da,Florea:2006si,Akerblom:2006hx,Cvetic:2007ku,Argurio:2007} that have focussed on ED2-branes at
D6-brane intersection. As stressed in \cite{Bianchi:2007fx}, one
immediate advantage of the viewpoint advocated here is the
consistency of the local description. Indeed, imposing twisted
tadpole cancellation \cite{Ibanez:1998qp,Bianchi:2000de} the models presented here and
all closely related settings of D-branes at singularities (not
necessarily of the $\Z_n$ kind) give rise to anomaly free
theories, while this is not necessarily the case for the `local'
models with intersecting D-branes. We can envisage the possibility
of extending our analysis to other $\Z_n$ singularities
\cite{Aldazabal:2000sa,Buican:2006sn} or even to Gepner models
\cite{Angelantonj:1996mw,Dijkstra:2004cc,Anastasopoulos:2006da} where many
if not all ingredients, such as the brane actions from gauge
kinetic functions including one-loop threshold effects
\cite{Antoniadis:1999ge,Lust:2003ky,Bianchi:2005sa,Anastasopoulos:2006hn}, are available.

In the present paper we have not addressed phenomenological
implications of the stringy instanton effects we have analyzed in
detail. We hope to be able to investigate these issues in this or
similar contexts with D-branes at singularities, where the
rigidity of the cycles is well understood and allows for the
correct number of fermionic zero-modes. Clearly additional (closed
string) fluxes neeeded for moduli stabilization \cite{Lust:2006zg,Lust:2006zh}
may change some of our present conclusions.

\section*{Acknowledgments}
 It is a pleasure to thank P.
Anastasopoulos, R. Argurio, C. Bachas, M. Bertolini, M. Billo, G. Ferretti,
M.L. Frau, A. Kumar, E. Kiritsis, I. Klebanov, S. Kovacs, A. Lerda, L.
Martucci, I. Pesandro, R. Russo and M. Wijnholt for valuable discussions.
Special thanks go to G. Pradisi for collaboration on
the computation of the string amplitudes and useful exchanges.
During completion of this work, M.B. was visiting the Galileo Galilei
Institute in Arcetri (FI) and thanks INFN for hospitality and
support. M.B. is very grateful to the organizers and participants
to the workshop ``String and M theory approaches to particle
physics and cosmology'' for creating a very stimulating
atmosphere.  This work was supported
in part by the CNRS PICS no. 2530 and 3059, INTAS grant 03-516346,
MIUR-COFIN 2003-023852, NATO PST.CLG.978785, the RTN grants
MRTNCT- 2004-503369, EU MRTN-CT-2004-512194, MRTN-CT-2004-005104
and by a European Union Excellence Grant, MEXT-CT-2003-509661.

%\bibliographystyle{JHEP}
%\bibliography{refZ3}

\providecommand{\href}[2]{#2}\begingroup\raggedright\endgroup

\end{document}